\documentclass[aps,pre,showpacs,preprint,floatfix]{revtex4}
\usepackage{epsf,amsmath,amssymb}
\newcommand{\be}{\begin{equation}}
\newcommand{\ee}{\end{equation}}
\newcommand{\bfi}{\begin{figure}}
\newcommand{\efi}{\end{figure}}
\begin{document}
\title{Nonlocal evolution of weighted scale-free networks}
\author{K.-I. Goh, B. Kahng, and D. Kim}
\affiliation{School of Physics and Center for Theoretical Physics,\\
Seoul National University NS50, Seoul 151-747, Korea}
\date{\today}
\begin{abstract}
We introduce the notion of globally updating evolution for a class
of weighted networks, in which the weight of a link is
characterized by the amount of data packet transport flowing
through it. By noting that the packet transport over the network
is determined nonlocally, this approach can explain
the generic nonlinear scaling between the strength and the degree
of a node. We demonstrate by a simple model that the
strength-driven evolution scheme recently introduced can be
generalized to a nonlinear preferential attachment rule,
generating the power-law behaviors in degree and in strength
simultaneously.
\end{abstract}
\maketitle

Network research has arisen as the interdisciplinary
subject for studying complex systems~\cite{rmp,portobook,siam,
vespigbook,networkbiology,koonin-book,watts}.
Although the binary (on/off) picture of connectivity
has been shown to be quite informative and led to
a number of important progress in our understanding of complex systems,
such as the ubiquity of power laws in its connectivity pattern,
the degree distribution \cite{ba}, $p_d(k)\sim k^{-\gamma}$,
one may need to put a step forward to describe them more realistically.
The weighted network, in which links between nodes in the network,
or nodes themselves, bear non-uniform weights, is one of the most
straightforward generalization in
this direction~\cite{yook,noh,braunstein,china,li-chen,barrat03,almaas,
barrat04,macdonald,akr,kpark,ak,dm}.
The weight may represent the intimacy between individuals
in social networks, or the bandwidths of routers and
optical cables in the Internet. The weighted network can be
characterized by the generalized adjacency matrix ${\bf W}$ whose
element $\{w_{ij}\}$ denotes the weight of the link connecting the node
$i$ to $j$. By definition, $w_{ij}=0$ when there is no link between $i$
and $j$. We will restrict our interest here to the case of non-negative
weight, $\{w_{ij}\ge0\}$ for all $(i,j)$ \cite{newman04,kertesz}.
The strength $s$ of a node \cite{barrat03}, given by
\be s_i = \sum_{j=1}^{N}w_{ij}, \ee
generalizes the concept of the degree, the number of links it has
in binary networks.
In terms of $\{w_{ij}\}$, the degree $k$ of a node may be written
as \be k_i = \sum_{j=1}^{N} \textrm{sgn}(w_{ij}). \ee

The analysis of weighted networks has been hindered mostly by the lack
of large-scale data for the real networks.
Recently, Barrat {\it et al.}~\cite{barrat03}
made the first detailed comparative
analysis on the structure of weighted networks
of the real world, the scientific coauthorship network (SCN) and the
world-wide airport network (WAN).
For the SCN, the weight of a link between two scientists
is given roughly by the frequency of their collaboration, the effective
number of papers they cowrote.
For the WAN, the weight is taken as the total number of passengers
of the direct flights between two connected cities.
Interestingly, the two networks reveal qualitatively different
organization of weight and network topology.
For the SCN, the strength of a node (scientist)
scales linearly with the degree,
that is, \be s(k) \sim k~.\label{eq:scaling1}\ee
On the other hand, the WAN exhibits a nonlinear scaling as
\be
s(k) \sim k^{\beta},
\label{eq:scaling2}
\ee
with $\beta\approx 1.5$.

Later, Barrat, Barthelemy, and Vespignani (BBV) introduced a
simple evolution model for such weighted networks~\cite{barrat04}.
Basically, the BBV model is similar to the Barab\'asi-Albert (BA)
model of binary scale-free (SF) network \cite{ba} in spirit,
containing the growth and the preferential attachment (PA) as
basic ingredients. But the difference comes in the aspects that
(i) strength at each node and weight at each link are introduced
and (ii) the PA rule in the strength instead of the degree is applied. 
That is, the probability $\Pi_i$ that an
existing node $i$ will receive a connection from a
newly-introduced node is proportional to its strength, $\Pi_i \sim
s_i$. The strength of the target node subsequently increases by a
constant amount $\Delta$ and so does the weight of the links
incident upon the target node in a linear fashion. Then the degree
and the strength scales linearly with each other and the
distributions of them both follow power laws. On the other hand,
the nonlinear scaling Eq.~(\ref{eq:scaling2}) observed in the WAN
requires different approaches. BBV proposed the nonlinear coupling
between node-strength and link-weight, but a finite cutoff in
node-strength was necessary to achieve the SF behavior in both the
degree and the strength distribution~\cite{barrat-model}.

In this Brief Report, we propose a packet transport driven
evolution model of weighted SF network, which can illustrate
the nonlinear relationship Eq.~(4). 
We first point out that
the weight used for the WAN in Ref.~\cite{barrat03} is actually
the amount of traffic through the link. From this perspective, the
evolution of such weighted network should be viewed as packet
transport driven. The distinguishing point here is that the
traffic flowing through the network is determined in a nonlocal
manner, in high contrast to the local weight evolution rule as
formulated in the BBV model \cite{barrat04}. A measure of such
traffic over the (binary) network is the quantity called the load
\cite{load}, or the betweenness centrality \cite{freeman}. The
load of a node, the vertex-load, is defined by the effective
number of data units passing through that node when every pair of
nodes in the network sends and receives a unit data in unit time
step. The data are assumed to be delivered only along the shortest
path(s) between the pair. When the data encounters a branching
point during the transport, they are supposed to be divided evenly
by the number of branches. Thus the load quantifies the level of
traffic, albeit in an idealized way. One can also define the
link-load in a similar fashion, as the effective number of data
units passing through a given link. For SF networks, the load of
each node is heterogeneous, and its distribution follows a power
law, $p_{L}(\ell)\sim \ell^{-\delta}$ \cite{load,pnas}. If the
ranks of a node in degree and in load are preserved, then the
scaling relation, \be \ell_i \sim k_i^{\eta}, \label{bketa} \ee
holds for each node $i$, where $\eta=(\gamma-1)/(\delta-1)$. This
relation is valid in the BA-type model and the Internet
\cite{bc-corr}.

\bfi[t] \centerline{\epsfxsize=\linewidth
\epsfbox{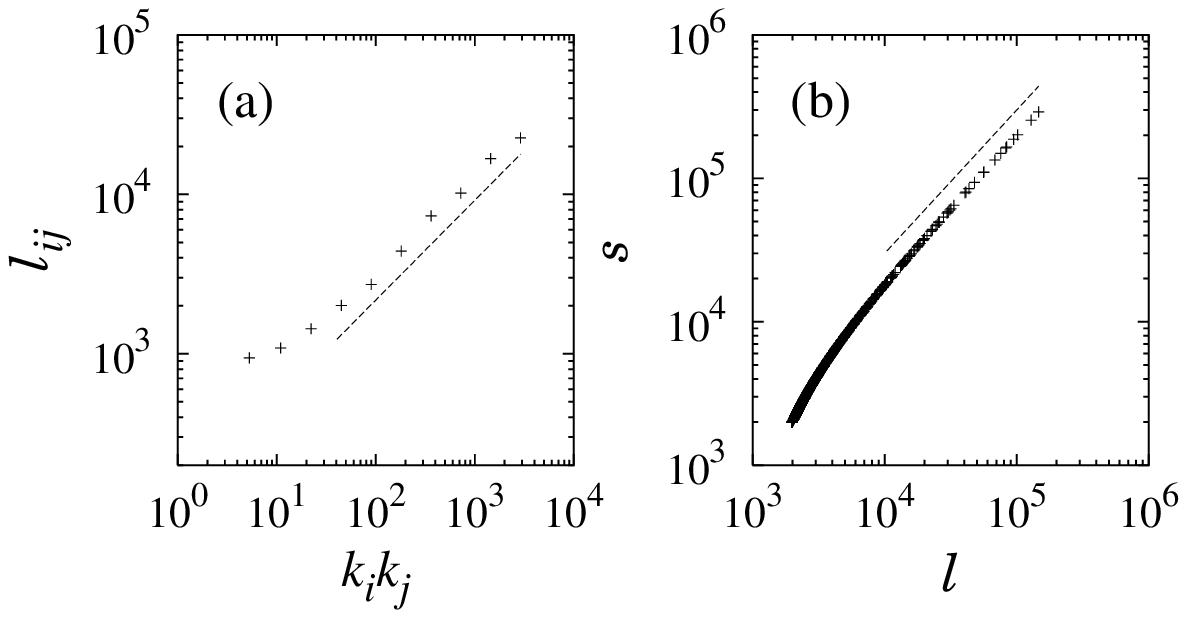}} \caption{(a) The link-load $\ell_{ij}$ vs.\
the product of the two degrees $k_ik_j$ of the nodes at each end
of the link in the BA model with the system size $N=10^3$. Data
points are logarithmically binned to reduce fluctuation. The
straight line with slope $0.63$ is drawn for reference. (b) The
relation between the node strength (the sum of the link-loads
attached to the node) and the load of the node. As indicated by
the guideline with slope $1$, they scale linearly for large
$\ell$. } \efi 
The weight of a link $w_{ij}$ and the product of the degrees of
the nodes at its ends $k_ik_j$ are found to be related as a power
law, $w_{ij}\sim (k_ik_j)^{\theta}$, and $\theta \approx 1/2$ for
the WAN \cite{barrat03}. Interestingly, the same half-power
scaling was observed in the metabolic network of {\em Escherichia
coli}, between the metabolic flux of a reaction and the degrees of
the participating metabolites~\cite{macdonald}. It is worthwhile
to note that such a correlation has also been observed in the relation of
the link-load $\ell_{ij}$ versus $k_ik_j$ for binary networks
\cite{han,macdonald}. We show in Fig.~1(a) such a relation for the
BA model~\cite{ba}, finding a slightly different scaling exponent
$\theta\approx 0.63$. 
So we may regard the link-load in the binary
network as an approximation of the weight in the weighted networks
characterized by the traffic level, such as the WAN. Although
actual traffic level may well depend also on more complicated
factors such as the geographic distances between the airports and
the queuing and transit delays at the airports, we exploit this
idea as a starting point for further discussion below. In this
setting, the strength of a node is given approximately by the load
of the node itself, i.e., \be s_i = \sum_{j \ne i}\ell_{ij} \sim
\ell_{i}, \label{eq:sl} \ee since the vertex-load of a node is
roughly one half of the sum of the link-loads of the links
connected to that node for large $\ell$ as shown in Fig.~1(b).

\bfi
\centerline{\epsfxsize=\linewidth \epsfbox{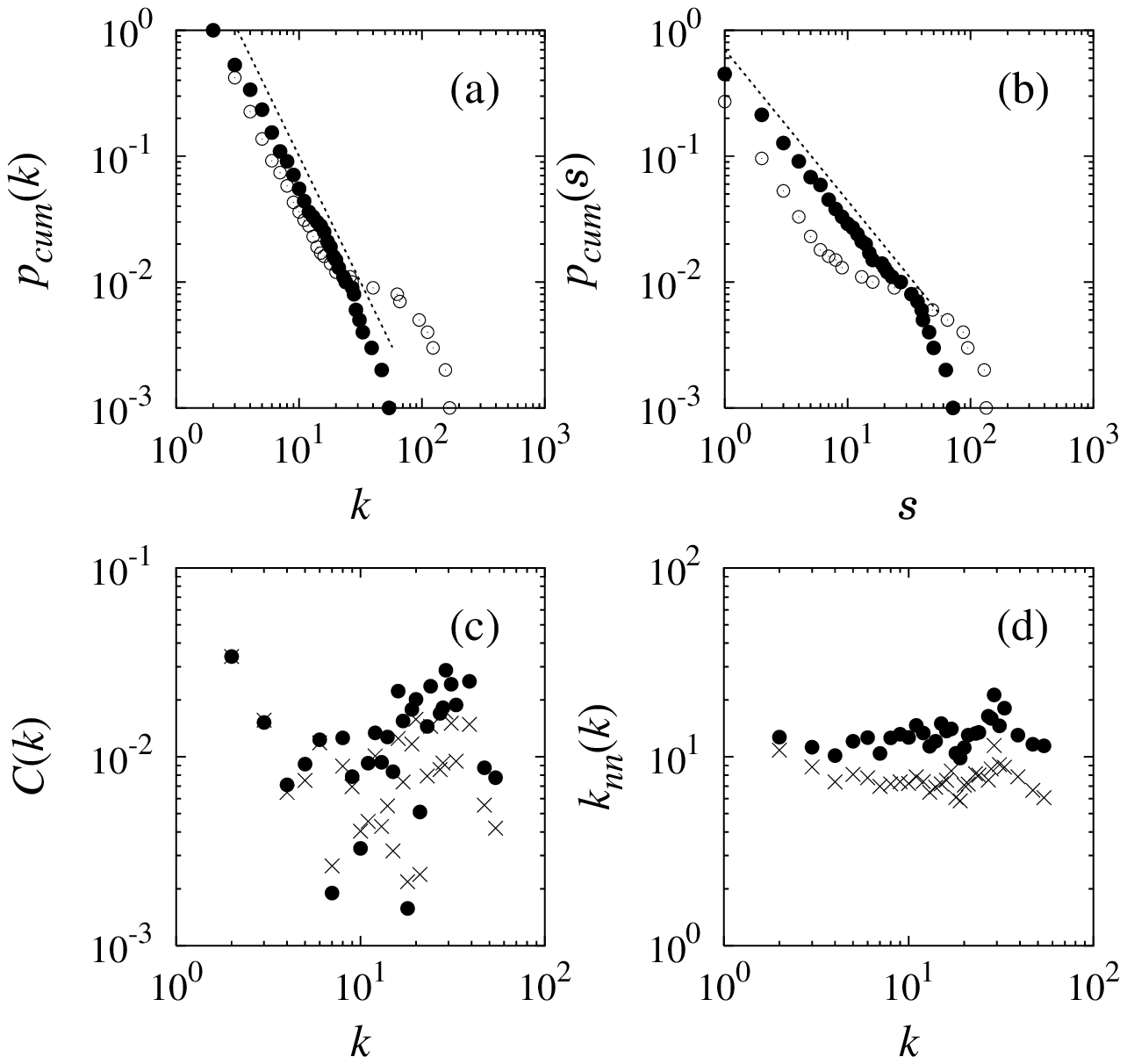}}
\caption{The cumulative distributions for the degree (a) and the strength (b)
of the packet transport-driven network growth with Eq.~(7).
The open circles denote the data for $\alpha=1$ and the filled
circles for $\alpha=0.6$. The guidelines have slopes $-2.0$ (a)
and $-1.2$ (b).
Note the humps for large degrees and large strengths in the case
of $\alpha=1$, indicating the breakdown of the SF nature.
The local clustering function $C(k)$ (c) and the average neighbor degree
function $k_{nn}(k)$ (d) for the network with $\alpha=0.6$.
The filled circles denote the weighted version of the corresponding
quantities introduced in Ref.~\cite{barrat03} and the cross symbols
the original binary version of them.
}
\efi

When the relation (\ref{bketa}) holds, the strength and the degree
of a given node scale nonlinearly as $s_i\sim k_i^{\eta}$.
Then the application of BBV scheme leads to $\Pi_i\sim s_i \sim k_i^{\eta}$
with $\eta > 1$ in most cases.
Such a super-PA in degree breaks the power-law
degree distribution~\cite{bu} as well as Eq.~(\ref{bketa}) itself.
To obtain an SF network, one must use the sub-PA
in strength,
\be
\Pi_i\sim s_i^{\alpha}\sim \ell_i^{\alpha}
\label{eq:pa}
\ee
with $\alpha=1/\eta$.
At a first glance, one may think that this rule of sub-PA in
strength would not generate a SF strength distribution, however,
we can achieve a SF strength distribution with the strength
being the load.

To demonstrate the above idea,
we simulate a growing network as follows.
(i) Initially, $m_0=3$ nodes are introduced and they are fully connected,
and we calculate the load at each node.
(ii) At each time step, a new node is added, and
attaches $m=2$ links to existing nodes selected
following the PA rule $\Pi_i\sim \ell_i^{\alpha}$.
(iii) The load of each node is recalculated.
This process is repeated $N$ times.
Figs.~2(a-b) show the simulation results, the degree and the strength (or load)
distributions of the packet transport-driven growth network for the cases
of $\alpha_1=1$ and $\alpha_2=1/\eta\approx0.6$.
Indeed we can see that for the $\alpha=1$ case, the power-law behaviors
of the degree and the strength distributions break down,
while for $\alpha=0.6$, the model reproduces both the power laws
with the degree exponent $\gamma \approx 3$ and the load exponent
$\delta \approx 2.2$, equivalent to those of the BA model.
These measured values of $\gamma$ and $\delta$ are consistent
with $\alpha=0.6$ through the relation $\alpha=(\delta-1)/(\gamma-1)$.
Other structural properties such as the clustering and the degree
mixing are also similar to those of the BA model [Figs.~2(c-d)].
It is of no surprise because this model with $\alpha=1/\eta$ is nothing
but the BA model rephrased in terms of the strength-driven evolution.

Note that in our model, the link weight
update is carried out through the change of the load,
which occurs by global reorganization of shortest pathways within the network.
As pointed out earlier, the load accounts for the traffic level
only in an approximated manner. The general framework described in
the paper, however, does not depend on what to use as the
traffic measure as long as the scaling relation Eq.(\ref{eq:scaling2})
holds for that quantity.

Surely, it is not always the case that the weight of a link is determined
by the level of traffic. In the SCN, for example, the weight measures
the direct affinity between the scientists, which is primarily determined
by their own attributes which is local in character.
Indeed, for SCN we can see no appreciable correlation between the
weight and the link-load of a link, as shown in Fig.~\ref{fig:condmat}.
In such a case, the BBV scheme could be more suitable than ours.
Thus in the weighted network modeling, one has first to discriminate
properly what the nature of the weight is in the system one wants
to describe.

Finally, we like to emphasize again the difference between the
strength used here and that of the BBV model. We also compare them 
with the vertex-load directly measured. For the purpose, we measure
the strength $s_i$ of node $i$ based on Eq.~(1) where the weight
$w_{ij}$ is given as the link-load $\ell_{ij}$, that is, 
$s_i=\sum_j \ell_{ij}$, and compare it with the vertex-load $\ell_i$ 
measured directly from the BBV network and the quantity $s_i^{({\rm
BBV})}$ defined in the BBV model~\cite{barrat04}. We note that
$s_i$ and $\ell_i$ ($s_i^{({\rm BBV})}$) are updated globally 
(locally) when a new node is added in the system. 
As shown in Fig.~\ref{fig:bbv}, we can find the scaling behaviors 
of $s\sim \ell \sim k^{1.33}$, which is consistent with the one 
obtained from the formula, $\ell \sim k^{(\gamma-1)/(\delta-1)}$, 
and $s^{({\rm BBV})}\sim k$. Those results support our claim.

\bfi[t] 
\centerline{\epsfxsize=0.8\linewidth \epsfbox{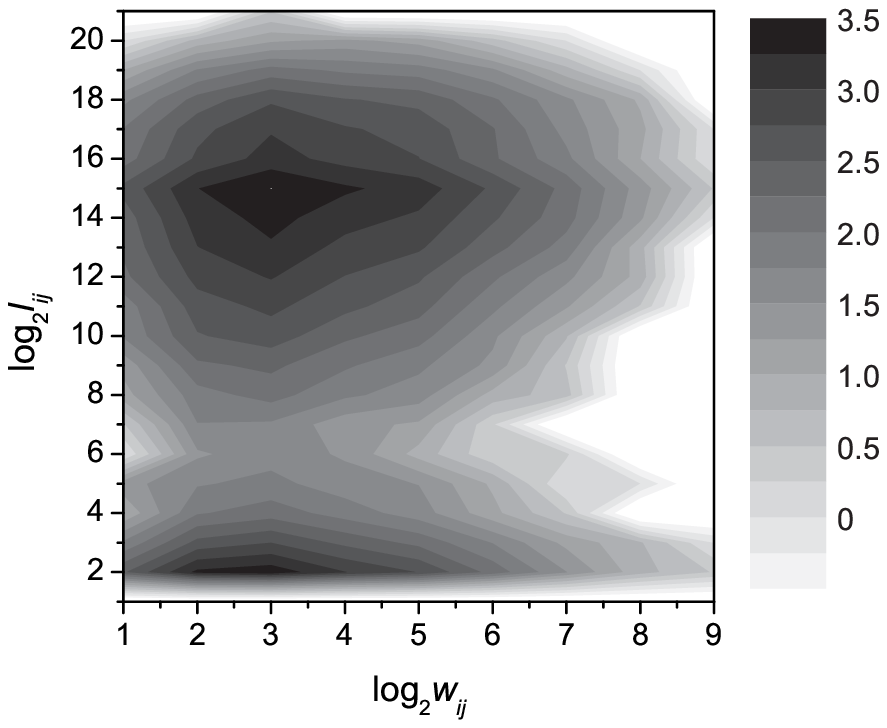}}
\caption{Histogram of the links having the link load $\ell_{ij}$
and the link weight $w_{ij}$ in the coauthorship network of
cond-mat subset of arXiv.org \cite{newman-pnas}. The grey level
denotes the number of links in each bin, in logarithm with base $10$. }
\label{fig:condmat} 
\efi 
\bfi[t]
\centerline{\epsfxsize=0.85\linewidth \epsfbox{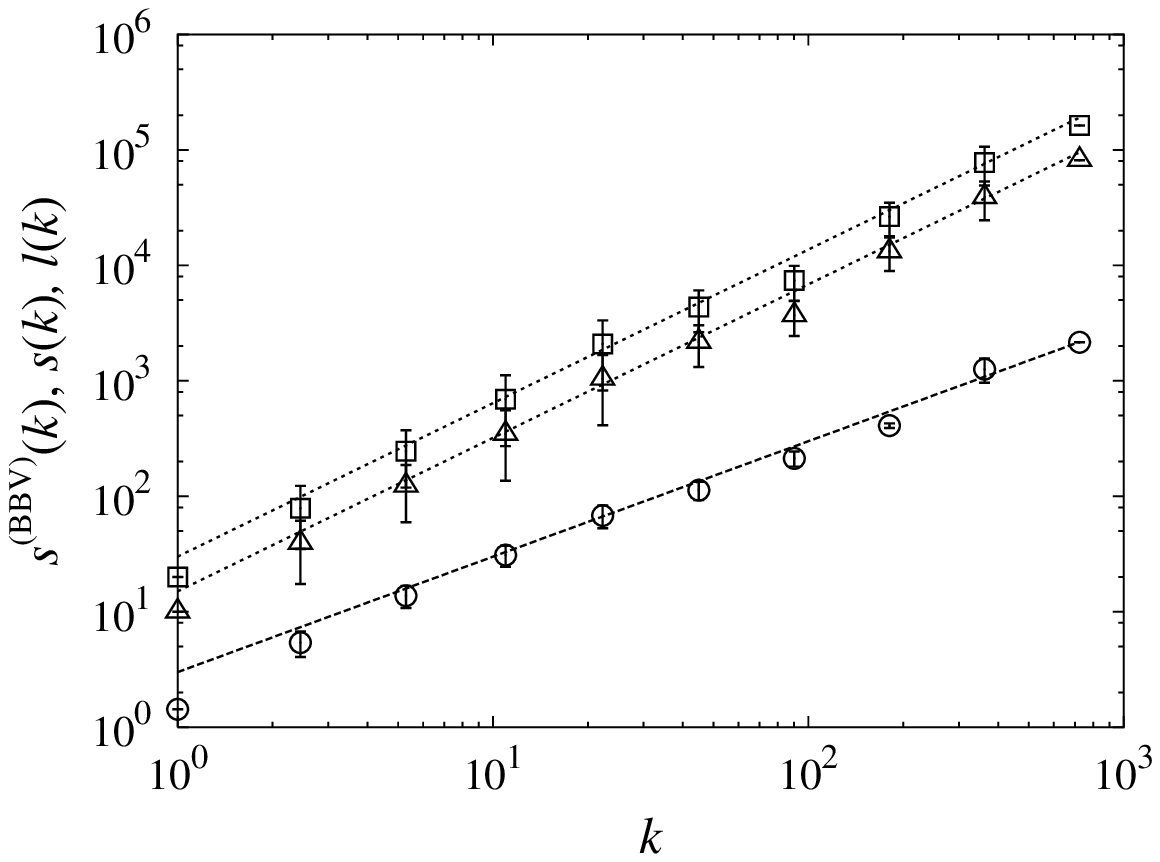}}
\caption{The strength $s(k)$ based on the link-load ($\Box$), 
the load $\ell(k)$ based on the vertex-load ($\triangle$), 
and the BBV strength $s^{({\rm BBV})}(k)$ ($\circ$) 
versus the degree $k$ for the BBV model network with system
size $N=10^4$, the number of links emanating from a new node
$m=1$, and the weight increment $w_0=\Delta=1.0$. The slopes of
the guide lines are 1.33 (dotted) and 1.0 (dashed), respectively.}
\label{fig:bbv} 
\efi

To conclude, we have introduced the notion of globally updating
evolution in a class of weighted networks, in which the weight is
characterized by the level of traffic flowing through the links.
The link-load is used as a measure of the traffic in our model.
This model explains the generic nonlinear scaling between the
strength (traffic) and the degree, observed in, e.g., the WAN. We
have also shown that the generalization of strength-driven
evolution into the nonlinear PA in strength is necessary to
produce a network which is SF both in degree and in strength.

{\sl Note in added}- After the first draft of the manuscript has been
completed, we have learned of a recent work by Bianconi~\cite{ginestra}
which introduced a model generating the nonlinear scaling relation
between strength and degree, Eq.(\ref{eq:scaling2}).
The model succeeds to obtain the nonlinear relationship by decoupling
the evolution of the network topology (degree) and the weights.
Although that model seems to be different from our conceptual
model at first, the two models share important aspects. First
its binary network structure grows following the BA model
and second, the weights are updated nonlocally rather than
locally as was assumed in the BBV model. Moreover,
Guimer\`a and Amaral~\cite{amaral} introduced a model to
illustrate the traffic in the WAN. The model also contains
the rule of adding internal links between existing nodes,
which are chosen by the PA rule combined with the ingredient of
reducing traveling length. Such a rule is applied to all nodes
in the network, not limited to neighbors of a newly added node,
which needless to say accounts for global updating of traffic
pathways. Such global reorganization of weights produces the
non-linear relationship between strength and degree, which
is the main conclusion of the work.

\begin{acknowledgments}
We would like to thank Mark Newman for making the arXiv.org coauthorship
network data available. This work is supported by the KOSEF Grant
No. R14-2002-059-01000-0 in the ABRL program.
\end{acknowledgments}


\begin{thebibliography}{99}
\bibitem{rmp} R. Albert and A.-L. Barab\'asi, Rev. Mod. Phys. {\bf 74}, 47 (2002).
\bibitem{portobook} S.~N. Dorogovtsev and J.~F.~F. Mendes, {\it Evolution of networks} (Oxford University Press, Oxford, 2003).
\bibitem{siam} M.~E.~J. Newman, SIAM Rev. {\bf 45}, 167 (2003).
\bibitem{vespigbook} R. Pastor-Satorras and A. Vespignani, {\it Evolution and Structure of Internet} (Cambridge University Press, Cambridge, 2004).
\bibitem{networkbiology} A.-L. Barab\'asi and Z. N. Oltvai, Nat. Genet. {\bf 5}, 101 (2004).
\bibitem{koonin-book} E. V. Koonin, Y. I. Wolf, and G. P. Karev (eds.), {\it Power Laws, Scale-free Networks and Genome Biology} (Landes Bioscience, Georgetown, 2004).
\bibitem{watts} D. J. Watts, Annu. Rev. Sociol. {\bf 30}, 243 (2004).
\bibitem{ba} A.-L. Barab\'asi and R. Albert, Science {\bf 286}, 509 (1999).

\bibitem{yook} S.~H. Yook, H. Jeong, A.-L. Barab\'asi, and Y. Tu, Phys. Rev. Lett. {\bf 86}, 5835 (2001).
\bibitem{noh} J.~D. Noh and H. Rieger, Phys. Rev. E {\bf 66}, 066127 (2002).
\bibitem{braunstein} L. A. Braunstein, S. V. Buldyrev, R. Cohen, S. Havlin, and H. E. Stanley, Phys. Rev. Lett. {\bf 91}, 168701 (2003).
\bibitem{barrat03} A. Barrat, M. Barthelemy, R. Pastor-Satorras, and A. Vespignani, Proc. Natl. Acad. Sci. USA \textbf{101}, 3747 (2004).
\bibitem{almaas} E. Almaas, B. Kovacs, T. Vicsek, Z. N. Oltvai, and A.-L. Barab\'asi, Nature {\bf 427}, 839 (2004).
\bibitem{barrat04} A. Barrat, M. Barthelemy, and A. Vespignani, Phys. Rev. Lett. {\bf 92}, 228701 (2004).
\bibitem{china} W. Li and X. Cai, Phys. Rev. E {\bf 69}, 046106 (2004).
\bibitem{kpark} K. Park, Y.-C. Lai, and N. Ye, Phys. Rev. E {\bf 70}, 026109 (2004)
\bibitem{li-chen} C. Li and G. Chen, cond-mat/0311333 (2003).
\bibitem{macdonald} P. J. Macdonald, E. Almaas, and A.-L. Barab\'asi, cond-mat/0405688 (2004).
\bibitem{akr} E. Almaas, P. L. Krapivsky, and S. Redner, cond-mat/0408275 (2004).
\bibitem{ak} T. Antal and P. L. Krapivsky, cond-mat/0408285 (2004).
\bibitem{dm} S.~N.~Dorogovtsev and J.~F.~F.~Mendes, cond-mat/0408343 (2004).
\bibitem{newman04} M. E. J. Newman, Phys. Rev. E {\bf 70}, 056131 (2004).
\bibitem{kertesz} J.-P. Onnela, J. Saram\"aki, J. Kertesz, and K. Kaski, cond-mat/0408629 (2004).
\bibitem{barrat-model} A. Barrat, M. Barthelemy, and A. Vespignani, Phys. Rev. E {\bf 70}, 066149 (2004).
\bibitem{load} K.-I. Goh, B. Kahng, and D. Kim, Phys. Rev. Lett. {\bf 87}, 278701 (2001).
\bibitem{freeman} L. C. Freeman, Sociometry {\bf 40}, 35 (1977).
\bibitem{pnas} K.-I. Goh, E. Oh, H. Jeong, B. Kahng, and D. Kim, Proc. Natl. Acad. Sci. USA {\bf 99}, 12583 (2002).
\bibitem{bc-corr} K.-I. Goh, E. Oh, B. Kahng, and D. Kim, Phys. Rev. E {\bf 67}, 017101 (2003).
\bibitem{han} P. Holme, B. J. Kim, C. N. Yoon, and S. K. Han, Phys. Rev. E {\bf 65}, 056109 (2004).
\bibitem{bu} P. L. Krapivsky and S. Redner, Phys. Rev. E {\bf 63}, 066123 (2001).
\bibitem{newman-pnas} M. E. J. Newman, Proc. Natl. Acad. Sci. USA {\bf 98}, 404 (2001).
\bibitem{ginestra} G. Bianconi, cond-mat/0412399 (2004).
\bibitem{amaral} R. Guimer\`a and Amaral, Eur. Phys. J. B {\bf 38,}
381 (2004).
\end{thebibliography}
\end{document}